\documentclass[10pt,conference,twocolumn]{IEEEtran}
\usepackage[utf8]{inputenc}
\usepackage[noadjust]{cite}
\usepackage{graphicx}
\usepackage{amsmath}
\usepackage{amssymb}
\usepackage{mathtools}

\usepackage{caption}
\usepackage{subcaption}
\usepackage{todonotes}
\usepackage{siunitx}
\usepackage{nicefrac}
\usepackage[hidelinks]{hyperref}
\usepackage{algorithm}
\usepackage{algpseudocode}

\usepackage{tikz}
\usepackage{pgfplots}
\usepackage{datenumber}
\usepgfplotslibrary{units}
\usetikzlibrary{arrows}
\usetikzlibrary{shapes.geometric, arrows, fit}
\usetikzlibrary{automata,positioning}

\definecolor{Set1-7-1}{RGB}{228,26,28}
\definecolor{Set1-7-2}{RGB}{55,126,184}
\definecolor{Set1-7-3}{RGB}{77,175,74}
\definecolor{Set1-7-4}{RGB}{152,78,163}
\definecolor{Set1-7-5}{RGB}{255,127,0}
\definecolor{Set1-7-6}{RGB}{166,86,40}
\definecolor{Set1-7-7}{RGB}{0,0,0}

\definecolor{leman}{RGB}{0,167,159}
\definecolor{vertdeau}{RGB}{194,221,176}
\definecolor{perle}{RGB}{202,199,199}
\definecolor{canard}{RGB}{0,116,128}
\definecolor{groseille}{RGB}{181,31,31}
\definecolor{zinzolin}{RGB}{92,36,131}
\definecolor{carotte}{RGB}{236,102,8}
\definecolor{chartreuse}{RGB}{200,211,0}

\newcommand{\deadline}{2024-05-01}

\usepackage[calc]{datetime2}
\newcommand{\daystodeadline}[0]{
  \DTMsavedate{DeadlineDate}{\deadline}%
  \DTMsavedate{TodayDate}{\the\year-\the\month-\the\day}%
  \newcount\mycttoday%
  \newcount\myctdeadline%
  \newcount\diff%
  \DTMsaveddatetojulianday{DeadlineDate}{\myctdeadline}%
  \DTMsaveddatetojulianday{TodayDate}{\mycttoday}%
  \diff=\myctdeadline%
  \advance\diff-\mycttoday%
  \the\diff%
}
\begin{document}
\author{\IEEEauthorblockN{Joachim Tapparel\IEEEauthorrefmark{1} and Andreas Burg\IEEEauthorrefmark{1}\\%
\IEEEauthorblockA{\IEEEauthorrefmark{1}Telecommunication Circuits Laboratory, \'Ecole polytechnique fédérale de Lausanne (EPFL), Switzerland}}}
\title{LoRa Fine Synchronization with Two-Pass Time and Frequency Offset Estimation}
\maketitle
\medskip

\begin{abstract}  
LoRa is currently one of the most widely used low-power wide-area network~(LPWAN) technologies.
The physical layer leverages a chirp spread spectrum modulation to achieve long-range communication with low power consumption.
Synchronization at long distances is a challenging task as the spread signal can lie multiple orders of magnitude below the thermal noise floor.
Multiple research works have proposed synchronization algorithms for LoRa under different hardware impairments.
However, the impact of sampling frequency offset~(SFO) has mostly either been ignored or tracked only during the data phase, but it often harms synchronization.
In this work, we extend existing synchronization algorithms for LoRa to estimate and compensate SFO already in the preamble and show that this early compensation has a critical impact on the estimation of other impairments such as carrier frequency offset and sampling time offset.
Therefore it is critical to recover long-range signals.
\end{abstract} 
\section{Introduction}
LoRa is a low-power wide-area network~(LPWAN) technology that has become very popular in recent years.
Its physical layer is based on a chirp spread spectrum~(CSS) modulation that allows long-range communication with low power consumption~\cite{loraPatent}.
Multiple spreading factors~(SF) can be used to spread the LoRa symbol in time and provide a trade-off between data rate and range.
As the power of the spread signal can lie multiple orders of magnitude below the thermal noise floor, the synchronization of LoRa frames relies on a known preamble that is prepended to every transmission. 
Multiple works have proposed synchronization algorithms for LoRa under different hardware impairments~\cite{ghanaatian2019lora,marquet2020lorasync,bernier2020low,li2021nelora,xhonneux2022low}.
However, the impact of sampling frequency offset~(SFO) has mostly been relegated to a secondary issue that could either be neglected or simply derived after the synchronization stage from the carrier frequency offset~(CFO) estimate. 
While it is true that the impact of SFO on the preamble is negligible for small spreading factors, it becomes detrimental for larger ones that accumulate more error due to their longer symbol and therefore also longer preamble duration.
In this work, we show that the presence of SFO in the preamble has a critical impact on the estimation of other impairments such as CFO and sampling time offset~(STO) and can therefore not be neglected during synchronization.
We consequently extend existing synchronization algorithms for LoRa to also estimate and compensate the SFO already in the preamble using a two-stage process.
This refined estimation process allows to extend the range of reference clock frequency offsets that can be tolerated by the synchronization algorithm and is especially beneficial for larger spreading factors.
\subsubsection*{Contributions} 
We first evaluate the impact of the sampling frequency offset on the LoRa preamble and derive a low-complexity phase correction to compensate sampling offset without complex resampling.
We then propose a two-pass time and frequency offset estimation for LoRa that is resilient to sampling frequency offset.
Finally, we evaluate the performance of the proposed synchronization algorithm through Monte Carlo simulations.

\section{LoRa Physical Layer}
\subsection{LoRa Signal Model}
LoRa uses a chirp spread spectrum modulation that is defined by a chirp signal with a linear increase in frequency over time.
The chirp waveform is defined by two main parameters.
The spreading factor~(SF) determines the number of chips per symbol as $N=2^\text{SF}$ and can be chosen in the range $\text{SF}\in\{5,\dots,12\}$.
The bandwidth $B$ determines the frequency span of the chirp signal.
Each LoRa symbol contains SF bits of information and has a duration $T_s=\frac{2^\text{SF}}{B}$.
The baseband chirp starts at a frequency $f_0= \frac{sB}{N}{-}\frac{B}{2}$, where $s\in \mathcal{S} {=} \{0,\dots,N{-}1\}$ is the modulated symbol value, and increases linearly until it reaches a frequency $\frac{B}{2}$.
The frequency chirp then folds back to the lower frequency $-\frac{B}{2}$ and continues to increase until the end of the symbol duration.
A LoRa symbol with value $s$, sampled at a rate $f_s$, can be expressed as~\cite{afisiadis2019error,chiani2019lora,tapparel2024grlora}
\begin{equation}
  x_s[n] =  e^{j2\pi\left(\frac{n^2}{2N}\left(\frac{B}{f_s}\right)^2+\left(\frac{s}{N}-\frac{1}{2}-u[n-n_\text{fold}]\right)\left(\frac{B}{f_s}\right)n\right)},
  \label{eq:upchirp}
\end{equation}
where $n \in \left\{0,\dots,N\frac{f_s}{B}-1\right\}$, where $n_\text{fold} = \frac{f_s}{B}(N-s)$ corresponds to the index at which the chirp frequency folds from $\nicefrac{B}{2}$ back to $\nicefrac{-B}{2}$, and where $u[n]$ denotes the unit step function.

\subsection{Demodulation}
In the following discussion, we consider the reception of the $l$-th LoRa symbol $\mathbf{y}_l$ received over an additive white Gaussian noise~(AWGN) channel, given by
\begin{equation}
  \mathbf{y}_l = \mathbf{x}_{s_l}+\mathbf{z}_l,
\end{equation}
where $\mathbf{x}_{s_l} = [x_{s_l}[0] \dots x_{s_l}[N-1] ]$ and where each element of $\mathbf{z}_l$ follows the complex Gaussian noise $\mathcal{CN}(0,\sigma^2)$ with variance $\sigma^2$.
A LoRa chirp sampled at a rate $f_s = B$ can efficiently be demodulated in three steps~\cite{chiani2019lora}:
First, the received symbol is multiplied element-wise with a reference down-chirp, i.e., a chirp $\mathbf{x}_0^*$, where $\cdot^*$ denotes the complex conjugate operator.
Second, the correlation between the received signal and the different symbol waveforms is computed efficiently with the discrete Fourier transform of the dechirped signal as
\begin{equation}
  \mathbf{\tilde Y}_l =  \text{DFT}(\mathbf{y}_l \odot \mathbf{x}_0^*).
\end{equation}
Finally, an estimate of the symbol value $\hat s$ is obtained by selecting the index with the maximum absolute value of $\mathbf{\tilde Y}_l$ as
\begin{equation}
  \hat{s}_l = \arg\max_{k\in\mathcal{S}}\left(\left|\tilde Y_l[k]\right|\right).
\end{equation}

\section{Preamble Synchronization}
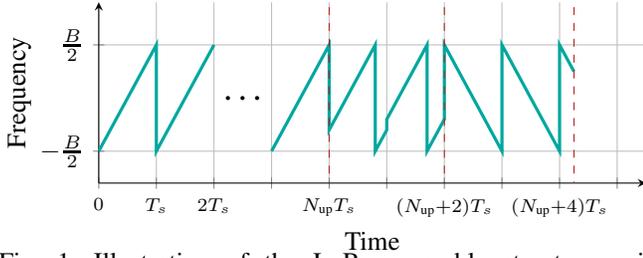
\begin{figure}[t]
  \begin{tikzpicture}
      \begin{axis}
          [
              width=\linewidth,
              height=0.45\linewidth,
              axis x line=bottom,
              axis y line=left,
              xlabel={Time},
              ylabel={Frequency},
              ylabel style={yshift=-5pt},
              xmin=0,
              xmax=9.5,
              ymin=-0.3,
              ymax=1.4,
              ytick={0,1},
              xtick={0,1,2,3,4,5,6,7,8,9},
              xticklabels={$0$,$T_s$,$2T_s$,,$N_\text{up}T_s$,,$(N_\text{up}{+}2)T_s$,,$(N_\text{up}{+}4)T_s$,,$N_\text{up}{+}7T_s$},
              yticklabels={$-\frac{B}{2}$,$\frac{B}{2}$},
              grid=major,
              xticklabel style={font=\scriptsize},
          ]
          \addplot[mark=none,very thick,leman] coordinates {
              (0,0) (1,1)                
              (1,0) (2,1)          
              };
          \addplot[mark=none,very thick,leman] coordinates {
            (3,0) (4,1) 
            (4,0.2) (4.8,1)
            (4.8,0) (5,0.2)  
            (5,0.3) (5.7,1)
            (5.7,0) (6,0.3)
            (6,1) (7,0)                
            (7,1) (8,0)                
            (8,1) (8.25,0.75)                
            };
            \node at (axis cs:2.55,0.5) {\Large \dots};

            \addplot[mark=none,dashed,groseille] coordinates {
              (4,-2) (4,2)                         
          };
          \addplot[mark=none,dashed,groseille] coordinates {
              (6,-2) (6,2)                         
          };
          \addplot[mark=none,dashed,groseille] coordinates {
              (8.25,-2) (8.25,2)                         
          };
      \end{axis}
  \end{tikzpicture}
\vspace{-10pt}
  \caption{Illustration of the LoRa preamble structure, with symbols of duration $T$ and chirp bandwidth $B$}  \label{fig:preamble}
\end{figure}

Every LoRa frame starts with a preamble that is used for synchronization.
The preamble structure is illustrated on \figurename~\ref{fig:preamble} and is composed of $N_\text{up}$ repetitions of the chirp $\mathbf{x}_0$, two modulated symbols called sync words, and two and a quarter down-chirps $\mathbf{x}^*_0$. 
The preamble can be used to estimate the two most critical impairments in LoRa synchronization, the carrier frequency offset~(CFO) and the sampling time offset~(STO).

The following sections are organized as follows: first we introduce the CFO and STO estimation methods used in state-of-the-art implementations. 
Second, we discuss the impact of the sampling frequency offset on the LoRa demodulation and synchronization.
Finally, we propose a two-pass time and frequency offset estimation for LoRa that is resilient to sampling frequency offset.

\subsection{Carrier Frequency Offset and Sampling Time Offset}

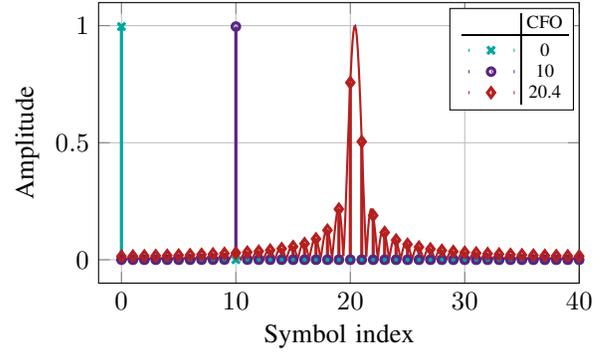
\begin{figure}[t]
  \centering
  \begin{tikzpicture}
    \begin{axis}[%
        name=main,
        width=.9\linewidth,
        height=.6\linewidth,
        xmin=-2,
        xmax=40,
        title style={font=\footnotesize},
        title style={at={(0.5,-0.3)}, anchor=north},
        xminorticks=true,
        xlabel={Symbol index},
        ylabel={Amplitude},
        xlabel style={yshift=3pt,},
        ylabel style={yshift=-8pt,},
        grid=both,
    ]
    \addplot [ycomb,color=leman,very thick,mark = x, mark size=2pt, opacity=1, domain = 0:64,samples=65] {abs(sin(pi*deg(x-0+0.001))/pi/(x-0+0.001))};
    \label{nocfo}
    

    \addplot [ycomb,color=zinzolin,very thick,mark = o, mark size=1.4pt, opacity=1, domain = 0:64,samples=65] {abs(sin(pi*deg(x-10+0.001))/pi/(x-10+0.001))};
    \label{cfo_int}

    \addplot [ycomb,color=groseille,very thick,mark = diamond, mark size=1.8pt, opacity=1, domain = 0:64,samples=65] {abs(sin(pi*deg(x-20.4))/pi/(x-20.4))};
    \label{cfo_int+frac}
    \addplot [color=groseille, opacity=.5,thick, domain = 0:64,samples=800] {abs(sin(pi*deg(x-20.4))/pi/(x-20.4))};


    \end{axis}
    \node [draw,fill=white,inner sep=2pt,at={(main.north east)},yshift=-2pt,xshift=-2pt,anchor=north east,]{
        \scriptsize
        \setlength{\tabcolsep}{2pt}
        \begin{tabular}{r|c}
            & CFO\\
            \hline
            \vspace{-6pt}\\
            \ref{nocfo} & 0  \\
            \ref{cfo_int} & 10  \\
            \ref{cfo_int+frac} & 20.4  \\
        \end{tabular}
    };
\end{tikzpicture}%
  \caption{Illustration of the impact of integer and fractional offset on the demodulated signal $\mathbf{\tilde Y}$}
  \label{fig:int_frac_offset_illust}
\end{figure}

As proposed in~\cite{bernier2020low,xhonneux2022low}, the CFO and STO can be separated into two parts, an integer part $L_x\in \mathbb{Z}$ and a fractional part $\lambda_x\in [-0.5,0.5)$ as
\begin{equation}
  \Delta f_c = \frac{B}{N}(L_\text{CFO}+\lambda_\text{CFO}), \quad \tau = \frac{L_\text{STO}+\lambda_\text{STO}}{B}.
\end{equation}
This representation of both offsets corresponds to the shift of the peak location in the demodulated signal $\mathbf{\tilde Y}$.
As illustrated in \figurename~\ref{fig:int_frac_offset_illust}, the integer offset shifts the peak location without changing the shape of the signal.
Conversely, the fractional offset changes the signal shape as the underlying cardinal sine function is not sampled at the time instant that result in a sampled dirac pulse. 
Due to their different impact on the demodulated signal, the integer and fractional offsets can be more easily estimated separately.
As discussed in~\cite{xhonneux2022low}, the order of the estimation of the different offset components should be performed as follow:
first, the fractional part of the CFO can be estimated based on the phase rotation between consecutive same-valued symbols using
\begin{equation}
  \hat\lambda_\text{CFO} = \frac{1}{2\pi}\angle\left(\sum_{l=2}^{N_\text{up}}\sum_{p=-2}^{2}  \tilde Y_l[{i{+}p}] \cdot \tilde Y_{l-1}[{i{+}p}]   \right),
  \label{eq:frac_cfo_est}
\end{equation}
where $i=\arg\max_k |\tilde Y_l[k] |$ and where $\angle(\cdot)$ denotes the argument of a complex number.
The estimated CFO can be compensated by adding a phase rotation to the received signal as 
\begin{equation}
  y'[n]= y[n]\cdot e^{-j2\pi \frac{\widehat{\text{CFO}}}{N}\frac{B}{f_s}n},\quad n\in\mathbb{N}_0.
  \vspace*{-2pt}
  \label{eq:cfo_comp}
\end{equation}
After the fractional CFO compensation, the fractional STO can be estimated using a pure tone frequency estimation method.
\textcolor{black}{Using the method proposed in~\cite{yang2011rctsl}, we can obtain an estimate of $\lambda_\text{STO}$ based on the energy of the three main spectral components of the dechirped signal as}
\begin{align}
  \mathbf{P} &= \sum_{l=1}^{N_\text{up}}\left| \text{DFT}_{2^ \text{SF}} (\mathbf{y}_l \odot \mathbf{x}_o^*) \right|^2, \\
  \hat\lambda_\text{STO} &= \frac{N}{2\pi} \frac{  P[i+1] -  P[i-1]}{u(  P[i+1]+  P[i-1] )+v  P[i] },\\
  u &= \frac{64\cdot N}{(\pi^5+32\pi)}, \quad v = \frac{u\pi^ 2}{4},
  \label{eq:frac_sto_est}
\end{align}
where $i=\arg\max_k(P[k]) $ and where DFT$_n$ denotes a DFT with a zero-padding of size $n$.
Finally, after resampling the signal to compensate the fractional STO, the integer parts of CFO and STO can be estimated using the up-chirps and down-chirps present in the preamble.
The demodulation of the up-chirps provides an estimated symbol value $\hat s_\text{up}$ while the demodulation of the down-chirps, using a reference waveform $\mathbf{x}_0$ for dechirping, provides an estimated symbol value $\hat s_\text{down}$.
Both values depend on the integer parts of the CFO and STO as
\begin{equation}
  \left\{
    \begin{array}{l@{\hspace{4pt}}l}
    \hat s_\text{up}&= (\hat L_\text{CFO} + \hat L_\text{STO})\bmod N,\\
    \hat s_\text{down}&= (\hat L_\text{CFO} - \hat L_\text{STO})\bmod N.
  \end{array}
  \right.
  \label{eq:up_down}
\end{equation}
Solving \eqref{eq:up_down} provides an estimate of the integer parts of the CFO and STO as
\begin{equation}
  \left\{
    \begin{array}{l@{\hspace{4pt}}l}
  \hat L_\text{CFO} &= \frac{1}{2}\Gamma_{N} \Big[ (\hat s_\text{up}+\hat s_\text{down}) \bmod N \Big],\\
  \vspace{-8pt}\\
  \hat L_\text{STO} &= (\hat s_\text{up}-\hat L_\text{CFO})\bmod N,
  \end{array}
  \right.
  \label{eq:int_cfo_sto_est}
\end{equation}
where
\begin{equation}
  \Gamma_N[k] = k-N\cdot u(k-\frac{N}{2}).
\end{equation}
We can note that only carrier frequency offsets values within $[\nicefrac{-B}{4},\nicefrac{B}{4}]$ can be estimated using this method as the solution of the system of equations is not unique.
However, the estimation range is sufficient for supporting the typical reference frequency offsets of low-cost radios.

\subsection{Sampling Frequency Offset}\label{sec:sfo}
\begin{figure}[t]
  \centering
  \begin{tikzpicture}

  \begin{axis}[%
    name=main,
    width=.8\linewidth,
    height=0.58\linewidth,
    ymode=log,
    xmin=0,
    xmax=41,
    ymin=0.99,
    ymax=1000,
    xminorticks=true,
    xlabel={Clock offset (ppm)},
    ylabel={\begin{tabular}{c}Maximum correctly\\ received bytes\end{tabular}},
    xticklabel shift={0pt},
    xlabel style={yshift=3pt,},
    xlabel style={yshift=0pt, },
    ylabel style={yshift=-3pt,},
    axis background/.style={fill=white},
    xmajorgrids,
    xminorgrids,
    ymajorgrids,
    yminorgrids,
  ]
  \addplot [color=chartreuse,  line width=1.5pt]
  table[row sep=crcr]{%
0.5	3906.25\\
1	1953.125\\
1.5	1302.08333333333\\
2	976.5625\\
2.5	781.25\\
3	651.041666666667\\
3.5	558.035714285714\\
4	488.28125\\
4.5	434.027777777778\\
5	390.625\\
5	390.625\\
7	279.017857142857\\
9	217.013888888889\\
11	177.556818181818\\
13	150.240384615385\\
15	130.208333333333\\
17	114.889705882353\\
19	102.796052631579\\
21	93.0059523809524\\
23	84.9184782608696\\
25	78.125\\
27	72.337962962963\\
29	67.3491379310345\\
31	63.0040322580645\\
33	59.1856060606061\\
35	55.8035714285714\\
37	52.7871621621622\\
39	50.0801282051282\\
41	47.6371951219512\\
};
\label{SF8}

\addplot [color=leman,  line width=1.5pt]
  table[row sep=crcr]{%
0.5	2197.265625\\
1	1098.6328125\\
1.5	732.421875\\
2	549.31640625\\
2.5	439.453125\\
3	366.2109375\\
3.5	313.895089285714\\
4	274.658203125\\
4.5	244.140625\\
5	219.7265625\\
5	219.7265625\\
7	156.947544642857\\
9	122.0703125\\
11	99.8757102272727\\
13	84.5102163461538\\
15	73.2421875\\
17	64.6254595588235\\
19	57.8227796052632\\
21	52.3158482142857\\
23	47.7666440217391\\
25	43.9453125\\
27	40.6901041666667\\
29	37.8838900862069\\
31	35.4397681451613\\
33	33.2919034090909\\
35	31.3895089285714\\
37	29.6927787162162\\
39	28.1700721153846\\
41	26.7959222560976\\
};
\label{SF9}

\addplot [color=zinzolin,  line width=1.5pt]
  table[row sep=crcr]{%
0.5	1220.703125\\
1	610.3515625\\
1.5	406.901041666667\\
2	305.17578125\\
2.5	244.140625\\
3	203.450520833333\\
3.5	174.386160714286\\
4	152.587890625\\
4.5	135.633680555556\\
5	122.0703125\\
5	122.0703125\\
7	87.1930803571429\\
9	67.8168402777778\\
11	55.4865056818182\\
13	46.9501201923077\\
15	40.6901041666667\\
17	35.9030330882353\\
19	32.1237664473684\\
21	29.0643601190476\\
23	26.5370244565217\\
25	24.4140625\\
27	22.6056134259259\\
29	21.0466056034483\\
31	19.6887600806452\\
33	18.4955018939394\\
35	17.4386160714286\\
37	16.4959881756757\\
39	15.6500400641026\\
41	14.8866234756098\\
};
\label{SF10}

\addplot [color=carotte,  line width=1.5pt]
  table[row sep=crcr]{%
0.5	671.38671875\\
1	335.693359375\\
1.5	223.795572916667\\
2	167.8466796875\\
2.5	134.27734375\\
3	111.897786458333\\
3.5	95.9123883928572\\
4	83.92333984375\\
4.5	74.5985243055556\\
5	67.138671875\\
5	67.138671875\\
7	47.9561941964286\\
9	37.2992621527778\\
11	30.517578125\\
13	25.8225661057692\\
15	22.3795572916667\\
17	19.7466681985294\\
19	17.6680715460526\\
21	15.9853980654762\\
23	14.595363451087\\
25	13.427734375\\
27	12.4330873842593\\
29	11.5756330818966\\
31	10.8288180443548\\
33	10.1725260416667\\
35	9.59123883928572\\
37	9.07279349662162\\
39	8.60752203525641\\
41	8.18764291158537\\
};
\label{SF11}

\addplot [color=groseille,  line width=1.5pt]
  table[row sep=crcr]{%
0.5	366.2109375\\
1	183.10546875\\
1.5	122.0703125\\
2	91.552734375\\
2.5	73.2421875\\
3	61.03515625\\
3.5	52.3158482142857\\
4	45.7763671875\\
4.5	40.6901041666667\\
5	36.62109375\\
5	36.62109375\\
7	26.1579241071429\\
9	20.3450520833333\\
11	16.6459517045455\\
13	14.0850360576923\\
15	12.20703125\\
17	10.7709099264706\\
19	9.63712993421053\\
21	8.71930803571428\\
23	7.96110733695652\\
25	7.32421875\\
27	6.78168402777778\\
29	6.31398168103448\\
31	5.90662802419355\\
33	5.54865056818182\\
35	5.23158482142857\\
37	4.9487964527027\\
39	4.69501201923077\\
41	4.46598704268293\\
};
\label{SF12}

\end{axis}
\node [draw,fill=white,inner sep=2pt,at={(main.north east)},yshift=5pt,xshift=2pt,anchor=north east ,]{
    \scriptsize
	\setlength{\tabcolsep}{2pt}
	\begin{tabular}{cc}
    \multicolumn{2}{c}{SF}\\
        \hline
        \vspace{-6pt}\\
        8& \ref{SF8} \\
        9&  \ref{SF9}\\
        10& \ref{SF10} \\
        11& \ref{SF11} \\
        12& \ref{SF12} \\
    \end{tabular}
};
\end{tikzpicture}%
  \caption{Number of bytes before systematic demodulation error for different reference clock offsets and spreading factors}
  \label{fig:sfo_impact}
\end{figure}
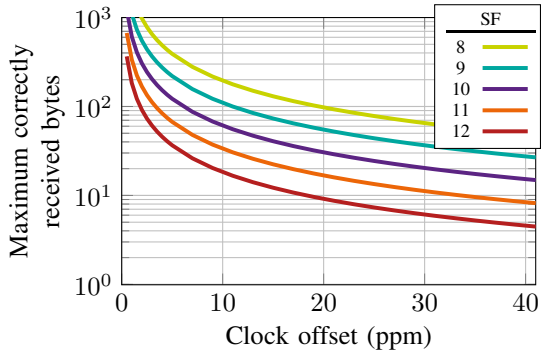

The mismatch in sampling rate between the transmitter and receiver causes the sampling time offset~(STO) to accumulate over time. 
Due to the accumulating offset, the location of the maximum peak of the demodulated signal $\mathbf{\tilde Y}$ drifts away from the correct symbol value.
\textcolor{black}{A reference frequency offset of $\Gamma$\,ppm causes an STO increase of $\gamma N$ per symbol, where $\gamma =\Gamma\times{10^{-6}} $.}
When the accumulated STO reaches a value of $0.5$, the demodulation will lead to a systematic error in the symbol estimation.
This drift limits the number of bytes that can be transmitted without error to $\left\lfloor \frac{1}{2\gamma N}\right\rfloor$.
\figurename~\ref{fig:sfo_impact} presents the number of bytes that can be received before reaching a systematic demodulation error for different reference clock frequency offsets and spreading factors.
While the impact of the SFO may be neglected for short frames with small spreading factors, it becomes critical for larger spreading factors.

During a frame, the accumulated STO can be compensated by periodically dropping or duplicating samples as soon as the accumulated time offset becomes larger than half the duration between two samples.
The period between sample drops (or duplications) can be derived from the accumulated STO as
\begin{equation}
  T_\text{comp} = \frac{f_s}{2 B \gamma}.
\end{equation}
Furthermore, the presence of SFO in the preamble impacts the estimation of other impairments such as CFO and STO.
The fractional STO estimation is the most impacted as the value changes during the preamble duration and, therefore, the pure tone frequency estimation method can not provide a valid estimation.
The fractional CFO estimation is also impacted as the phase rotation between consecutive symbols has an additional term due to the SFO.
The dechirped symbol $\tilde y_l[n] = y_l[n] x_0^*[n]$ impacted by both fractional CFO and SFO can be expressed as~\cite{ghanaatian2019lora}
\begin{equation}
  \tilde y_l[n] = e^{j2\pi n\left(\frac{s}{N}\frac{B}{f_s'}\right)}
  e^{j2\pi n\left[\frac{\lambda_\text{CFO}}{N}(n+lN)+l\left(\frac{B^2}{f_s'^2}-\frac{B}{f_s'}\right)  \right]}.
\end{equation}
After some algebraic manipulations, the phase difference between two consecutive symbols is found to be
\begin{equation}
  \angle (\tilde y_l[n]\tilde y^{(l-1)}[n]) = 2\pi\left[\lambda_\text{CFO} + \left(\frac{B^2}{f_s'^2}-\frac{B}{f_s'}\right)\right].
\end{equation}
Given that the fractional CFO estimation is using $f_s=B$, the phase difference can be written as a function of the clock offset $\gamma$ as 
\begin{equation}
  \angle (\tilde y_l[n]\tilde y^{(l-1)}[n]) = 2\pi\left[ \lambda_\text{CFO}-\frac{\gamma}{(\gamma+1)^2} \right].
  \label{eq:sfo_impact_frac_cfo}
\end{equation}
From~\eqref{eq:sfo_impact_frac_cfo}, we can notice that there is an error term in the fractional CFO estimation. 
However, the error term is much smaller than the fractional CFO itself as $\gamma \sim \SI{1e-5}{}$ and, therefore, can be neglected.
The estimation of both integer CFO and STO have a higher probability of having an error of one sample as the energy of the symbol is leaking in the directly adjacent bins of $\mathbf{\tilde Y}$, increasing the probability to obtain a symbol estimate off by $\pm 1$.
We should note that an $L_\text{CFO}$ estimate off by $\pm1$ only causes an error of frequency of $\pm \nicefrac{B}{2^\text{SF}}\,\si{\hertz}$, therefore, the larger the SF, the smaller the error.

\subsection{Two-Pass Time and Frequency Offset Estimation}
Using the fact that most IoT radios use the same reference oscillator for generating both the up- and down-conversion as well as the sampling clock, an estimation of the SFO can be obtained from the CFO estimate using
\begin{equation}
  \hat\gamma = (\hat L_\text{CFO}+ \hat \lambda_\text{CFO}) \frac{B}{N f_c}.
  \label{eq:sfo_from_cfo}
\end{equation}

As discussed in Section~\ref{sec:sfo}, the estimation of the CFO is only slightly degraded by the presence of an SFO.
Hence, we propose to perform a two-pass time and frequency offset estimation, where the first pass is used to obtain an estimation of the SFO and the second pass is used to refine the estimation of the CFO and STO after SFO compensation.

In order to compensate the SFO in the preamble, we propose to apply a sample-by-sample phase correction as a simple dropping or duplicating of samples would degrade the performance of the CFO and STO estimations.
This approach is motivated by the fact that the $l$-th LoRa symbol sampled at a frequency $f_s'$, introduced in \cite[Eq.~(11)]{ghanaatian2019lora}, can be simplified for the repeating preamble up-chirps as
\begin{equation}
  y_{l}[n]=e^{j2\pi\left[\frac{B^2}{2N}\left(\frac{n}{f_s'}+\Delta \tau_l\right)^2 -\frac{1}{2}B \left(\frac{n}{f_s'}+\Delta \tau_l\right) \right]},
  \label{eq:rec_sfo_upchirp}
\end{equation}
where $\Delta \tau_l = l N\left(\frac{f_s-f_s'}{Bf_s'}\right)$ corresponds to the accumulated time offset for the $l$-th symbol and where $f_s'=f_s(1+\gamma)$.
The phase of the received symbol can then be expressed as
\begin{align*}
  \angle(y_{l}[n])=\text{\footnotesize $2\pi\left[\frac{B^2}{2N }\left(\frac{n}{f_s'}\right)^2 +  \left(l \frac{B}{f_s'} \frac{f_s-f_s'}{f_s'} -\frac{B}{2f_s'}  \right) n \right]+\psi_l$},
   \raisetag{-5pt}
\end{align*}
where $\psi_l$ is a constant phase term that does not depend on $n$.
The additional phase rotation $\Delta\phi_l[n] \coloneqq \angle (y_{l}[n] x_0^*[n])$ introduced by the SFO is given by
\begin{align*}
  \Delta\phi_l[n]=\text{\footnotesize $2\pi\left[\frac{B^2}{2N}\frac{f_s^2{-}f_s'^2}{f_s^2f_s'^2}n^2+\left[l \left(\frac{B^2}{f_s'^2}{-}\frac{B}{f_s'}\right) -\frac{B}{2}\frac{f_s{-}f_s'}{f_s f_s'} \right] n \right]{+}\psi_l$}.
  \raisetag{-5pt}
\end{align*}
As we use a non-coherent demodulation and as an absolute phase offset does not impact the different CFO and STO estimation methods, we can ignore the constant phase term $\psi_l$.
Finally the additional phase term introduced by the SFO in the preamble is given by
\begin{align}
  \Delta\phi[n] = \text{\footnotesize $2\pi\left\{\frac{B^2}{2N}\frac{f_s^2{-}f_s'^2}{f_s^2f_s'^2}\bar{n}^2+\left[\left\lfloor{\frac{n}{N}}\right\rfloor \left(\frac{B^2}{f_s'^2}{-}\frac{B}{f_s'}\right) -\frac{B}{2}\frac{f_s{-}f_s'}{f_s f_s'} \right]\bar{n}\right\}$},
  \raisetag{-5pt}
  \label{eq:sfo_comp}
\end{align}
where $\bar{n} = n\bmod 2^\text{SF}$.
We should note that the phase term of the down-chirps is given by $-\Delta\phi[n]$, which can be derived similarly to the up-chirp case described previously.

\begin{figure}
  \centering
  \begin{tikzpicture}[every node/.style={text width=120pt,minimum height=2em,rectangle,font=\small, align=center, fill=perle!40, draw=black!60 }, node distance=8pt]
    \node[draw=none,fill=none,minimum height=1em] (rough) {Preamble};
    \node[draw, below=of rough] (sfo) {Compensate SFO using \eqref{eq:sfo_comp}};
    \node[draw,below=of sfo] (lcfo) {$\lambda_\text{CFO}$ estimation \eqref{eq:frac_cfo_est} \&\\ compensation \eqref{eq:cfo_comp}};
    \node[draw, below=of lcfo] (lsto) {$\lambda_\text{STO}$ estimation \eqref{eq:frac_sto_est} \&\\ compensation (resampling)};
    \node[ below=of lsto] (L) {Est. of $L_\text{STO}$ and $L_\text{CFO}$ \eqref{eq:int_cfo_sto_est} \&\\ comp. of CFO with \eqref{eq:cfo_comp} and STO with sample realignment};

    \node[ below=of L] (sfo_est) {SFO estimation using \eqref{eq:sfo_from_cfo}};
    \node[draw,diamond,aspect=3, below=of sfo_est, text width = 40pt,minimum height=0em,font=\footnotesize] (rhombus) {$\widehat{\text{SFO}} \cdot 2^\text{SF}{<}\theta$};

    \draw[->] (rough) -- (sfo);
    \draw[->] (sfo) -- (lcfo);
    \draw[->] (lcfo) -- (lsto);
    \draw[->] (lsto) -- (L);
    \draw[->] (L) -- (sfo_est);
    \draw[->] (sfo_est) -- (rhombus);
    \draw[->] (rhombus) --  ++ (0, -1)node[midway, right,draw=none, text width=20pt,fill=none] {Yes};
    \draw[->] (rhombus) -- ++ (2.9, 0) node[midway, below,draw=none, text width=20pt,fill=none] {No}|-(sfo.east);


    \node[shape=circle,draw=none,fill=groseille,text=white,inner sep=0pt,text width=3pt,minimum size=8pt, left=of sfo] (num1) {\scriptsize{1}};
    \node[shape=circle,draw=none,fill=groseille,text=white,inner sep=0pt,text width=3pt,minimum size=8pt, left=of lcfo] (num2) {\scriptsize 2};
    \node[shape=circle,draw=none,fill=groseille,text=white,inner sep=0pt,text width=3pt,minimum size=8pt, left=of lsto] (num3) {\scriptsize 3};
    \node[shape=circle,draw=none,fill=groseille,text=white,inner sep=0pt,text width=3pt,minimum size=8pt, left=of L] (num4) {\scriptsize 4};
    \node[shape=circle,draw=none,fill=groseille,text=white,inner sep=0pt,text width=3pt,minimum size=8pt, left=of sfo_est] (num5) {\scriptsize 5};
    
\end{tikzpicture}
  \caption{Proposed iterative time and frequency offset estimation for LoRa}
  \label{fig:sync_algo}
\end{figure}
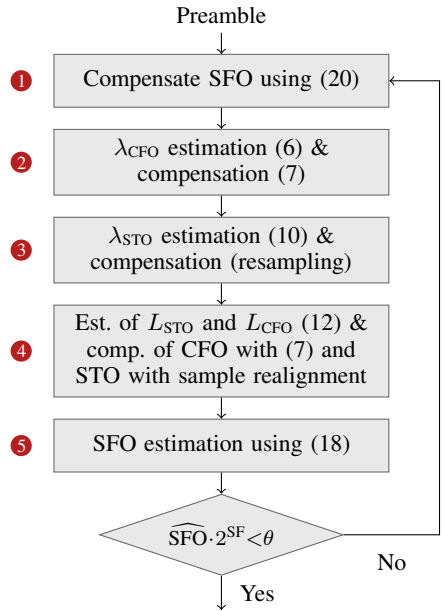

Using the previously described estimation and compensation methods, one iteration of the synchronization algorithm, illustrated in \figurename~\ref{fig:sync_algo}, can be summarized as follows:
\begin{enumerate}
  \item Compensation of the SFO in the preamble by removing the phase given in \eqref{eq:sfo_comp} from the received signal.
  \item Estimation and compensation of the fractional CFO using \eqref{eq:frac_cfo_est} and \eqref{eq:cfo_comp}, respectively.
  \item Estimation of the fractional STO using \eqref{eq:frac_sto_est} and compensation by resampling.
  \item Estimation of the integer CFO and STO using \eqref{eq:int_cfo_sto_est} and compensation of the integer CFO using \eqref{eq:cfo_comp} and of the integer STO by sample realignment.
  \item Estimation of the SFO from the CFO estimate using \eqref{eq:sfo_from_cfo}.
\end{enumerate}

The second iteration of the synchronization algorithm can be performed using the same steps as the first iteration, but with the SFO compensated in the preamble using the estimated value from the first iteration.
While the second pass greatly improves the synchronization performance for large spreading factors, additional iterations only provide marginal gains.
Furthermore, based on the value of the estimated SFO and the spreading factor used, one can decide to bypass the second iteration to save computational resources if the accumulated STO per symbol is smaller than a threshold $\theta$. 

\section{Results}\label{sec:results}
We now evaluate the performance of the proposed synchronization algorithm through Monte Carlo simulations.
In the following, we consider a LoRa signal that uses a bandwidth of \SI{250}{\kilo\hertz} and carrier frequency of \SI{868}{\mega\hertz}
We simulate transmission over an AWGN channel with a reference clock offset $\gamma$ that introduces a CFO $\Delta f_c= \gamma f_c$ and an SFO $\Delta f_s=\gamma B$.
Additionally, the results are averaged over random STO values, uniformly distributed in the range $[0,2^\text{SF})$.
\subsection{Offsets Estimation}
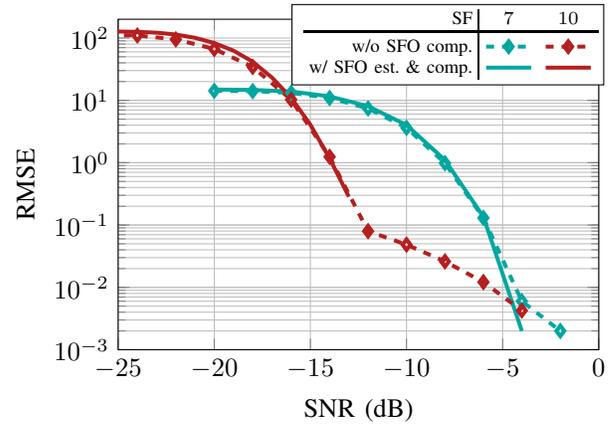
\begin{figure}
  \centering
  \begin{tikzpicture}

\begin{axis}[%
name=main,
width=.9\linewidth,
height=0.675\linewidth,
ymode=log,
xmin=-25,
xmax=0,
ymin=1e-3,
ymax=200,
xlabel={SNR (dB)},
ylabel={RMSE},
xticklabel shift={0pt},
xlabel style={yshift=0pt, },
axis background/.style={fill=white},
xmajorgrids,
xminorgrids,
ymajorgrids,
yminorgrids,
]

\addplot [color=leman, dashed,mark options={solid},  mark=diamond,line width=1.5pt]
  table[row sep=crcr]{%
-20	14.2111984012609\\
-18	14.0398514237153\\
-16	13.1991942178301\\
-14	10.9000166972349\\
-12	7.41062615438129\\
-10	3.62537225674827\\
-8	0.982230115604282\\
-6	0.130184484482599\\
-4	0.006\\
-2	0.002\\
0	0\\
2	0\\
4	0\\
6	0\\
8	0\\
10	0\\
};
\label{cfo_int:sf7_nc_it1}

\addplot [color=leman,line width=1.5pt]
  table[row sep=crcr]{%
-20	14.894413180787\\
-18	14.7084279241529\\
-16	13.8585479758884\\
-14	11.5954052969269\\
-12	8.04694675016556\\
-10	4.0255002173643\\
-8	1.09989454039922\\
-6	0.135321838592298\\
-4	0.002\\
-2	0\\
0	0\\
2	0\\
4	0\\
6	0\\
8	0\\
10	0\\
};
\label{cfo_int:sf7_nc_it2}

\addplot [color=groseille, dashed, mark options={solid},  mark=diamond, line width=1.5pt]
  table[row sep=crcr]{%
-28	113.531964714789\\
-26	113.012842518008\\
-24	109.800170464349\\
-22	94.9991713542807\\
-20	66.6372597125662\\
-18	34.3286316651276\\
-16	10.2583317357161\\
-14	1.24358272744518\\
-12	0.0797496081495075\\
-10	0.0484767985741633\\
-8	0.0260768096208106\\
-6	0.0121655250605964\\
-4	0.00424264068711928\\
-2	0\\
0	0\\
};
\label{cfo_int:sf10_nc_it1}

\addplot [color=groseille, mark=none, line width=1.5pt]
  table[row sep=crcr]{%
  -29	128.231094091878\\
-28	128.131173154701\\
-27	128.01809443981\\
-26	127.718412885535\\
-25	127.058122644717\\
-24	125.505741685391\\
-23	121.738906919686\\
-22	113.826047045481\\
-21	100.672839495069\\
-20	83.123918904248\\
-19	62.6465835301495\\
-18	41.6085030492567\\
-17	23.209867341284\\
-16	11.0076438895887\\
-15	4.06890820736964\\
-14	1.17386966908597\\
-13	0.284010563183837\\
-12	0\\
-11	0\\
-10	0\\
-9	0\\
-8	0\\
-7	0\\
-6	0\\
-5	0\\
-4	0\\
-3	0\\
-2	0\\
-1	0\\
0	0\\
1	0\\
};
\label{cfo_int:sf10_nc_it2}

\end{axis}
\node [draw,fill=white,inner sep=2pt,at={(main.north east)},yshift=-25pt,xshift=2pt,anchor=south east,]{
    \scriptsize
	\setlength{\tabcolsep}{2pt}
	\begin{tabular}{r|cc}
        SF	& $7$ &$10$  \\
        \hline	\vspace{-6pt}
        & & \\
        w/o SFO comp. &\ref{cfo_int:sf7_nc_it1} &\ref{cfo_int:sf10_nc_it1} \\
        w/ SFO est. \& comp. &\ref{cfo_int:sf7_nc_it2} &\ref{cfo_int:sf10_nc_it2}\\
    \end{tabular}
};
\end{tikzpicture}%
  \caption{RMSE of integer CFO estimate with a frequency offset $\gamma=\SI{40}{ppm}$ for different spreading factors}
  \label{fig:cfo_int_err}
\end{figure}

\begin{figure}
  \centering
%
%
\definecolor{mycolor1}{rgb}{0.00000,0.44706,0.74118}%
\definecolor{mycolor2}{rgb}{0.85098,0.32549,0.09804}%
\definecolor{mycolor3}{rgb}{0.92941,0.69412,0.12549}%
\definecolor{mycolor4}{rgb}{0.49412,0.18431,0.55686}%
\begin{tikzpicture}

\begin{axis}[%
name=main,
width=.9\linewidth,
height=0.675\linewidth,
ymode=log,
xmin=-28,
xmax=5,
ymin=1e-2,
ymax=0.5,
xminorticks=true,
xlabel={SNR (dB)},
ylabel={RMSE},
xticklabel shift={0pt},
xlabel style={yshift=0pt, },
axis background/.style={fill=white},
xmajorgrids,
xminorgrids,
ymajorgrids,
yminorgrids,
]

\addplot [color=leman, dashed, line width=1.5pt]
  table[row sep=crcr]{%
-20	0.279865189629111\\
-18	0.252242526140332\\
-16	0.193472056059919\\
-14	0.138347149360441\\
-12	0.0975143052750551\\
-10	0.069174836947763\\
-8	0.0549563803483726\\
-6	0.0486050331140115\\
-4	0.0449741506758448\\
-2	0.0425693772736489\\
0	0.0411056552384284\\
2	0.0402009220476494\\
4	0.0395826847487774\\
6	0.0390758306266539\\
8	0.0387267330620325\\
10	0.0384141903772145\\
};
\label{sf7_nc_it1}

\addplot [color=leman, line width=1.5pt]
  table[row sep=crcr]{%
-20	0.280295870164066\\
-18	0.253576633713852\\
-16	0.196282953518823\\
-14	0.141782953487896\\
-12	0.0986198016135191\\
-10	0.0662711813732612\\
-8	0.0491271184691057\\
-6	0.0418076044075239\\
-4	0.0373433811642944\\
-2	0.0342214899521488\\
0	0.0322081137886066\\
2	0.0309829963855247\\
4	0.0301454606420057\\
6	0.0296033105414079\\
8	0.0292453119084792\\
10	0.0290822200236229\\
};
\label{sf7_nc_it2}

\addplot [color=zinzolin, dashed, line width=1.5pt]
  table[row sep=crcr]{%
-23	0.283362397989089\\
-21	0.265531381149745\\
-19	0.213818952858711\\
-17	0.151585467321307\\
-15	0.109229034345843\\
-13	0.0831043465889576\\
-11	0.0711444292817005\\
-9	0.0660107932643109\\
-7	0.0629446813196507\\
-5	0.060750181100006\\
-3	0.0593485261221972\\
-1	0.0585727588240839\\
1	0.0580438803141071\\
3	0.0577129211516755\\
5	0.0575091170481467\\
7	0.0574380426406951\\
};
\label{sf8_nc_it1}

\addplot [color=zinzolin, line width=1.5pt]
  table[row sep=crcr]{%
-23	0.283448402357864\\
-21	0.266584522131418\\
-19	0.216734157628847\\
-17	0.153294393033405\\
-15	0.10517712930484\\
-13	0.0692359226774945\\
-11	0.0492776986979582\\
-9	0.0415546778944442\\
-7	0.0372541342412346\\
-5	0.0340606325940707\\
-3	0.0320636140302089\\
-1	0.0306689816278013\\
1	0.0297330435994896\\
3	0.0291290926778791\\
5	0.0288016264253393\\
7	0.0285696833904651\\
};
\label{sf8_nc_it2}

\addplot [color=carotte, dashed, line width=1.5pt]
  table[row sep=crcr]{%
-26	0.284167674720683\\
-24	0.267696608488201\\
-22	0.22240777541394\\
-20	0.171988876145472\\
-18	0.139450318843357\\
-16	0.120620589382313\\
-14	0.11384917596514\\
-12	0.110952007578663\\
-10	0.108945202377472\\
-8	0.107277965189091\\
-6	0.10618142686866\\
-4	0.105399095983721\\
-2	0.104944716331135\\
0	0.104449548528629\\
2	0.104144585178252\\
4	0.103929250733733\\
};
\label{sf9_nc_it1}

\addplot [color=carotte, line width=1.5pt]
  table[row sep=crcr]{%
-26	0.284717474787585\\
-24	0.269727455319783\\
-22	0.225540101574142\\
-20	0.168638620218828\\
-18	0.118745487729826\\
-16	0.0751303979028849\\
-14	0.0501111129342979\\
-12	0.0418548917557508\\
-10	0.0375217174880275\\
-8	0.0348500894492271\\
-6	0.0330685070391597\\
-4	0.0320400383301745\\
-2	0.0313839630044225\\
0	0.0310176739942474\\
2	0.0308011234035836\\
4	0.0306831443153105\\
};
\label{sf9_nc_it2}

\addplot [color=groseille, dashed, line width=1.5pt]
  table[row sep=crcr]{%
  -29	0.28708461037728\\
  -28	0.286427684671444\\
  -27	0.28445489696719\\
  -26	0.279639484664604\\
  -25	0.271380728717315\\
  -24	0.260629398358019\\
  -23	0.2497932605975\\
  -22	0.238739132653658\\
  -21	0.228183364604594\\
  -20	0.220001522292415\\
  -19	0.215852530101788\\
  -18	0.213984331868906\\
  -17	0.213765840129757\\
  -16	0.213596733923578\\
  -15	0.213132486057356\\
  -14	0.212660513401335\\
  -13	0.212140415536411\\
  -12	0.211863124163969\\
  -11	0.211614100570157\\
  -10	0.211405308650151\\
  -9	0.211331424574359\\
  -8	0.211295413301285\\
  -7	0.211296356785694\\
  -6	0.211376947210201\\
  -5	0.211398730989078\\
  -4	0.21141393558975\\
  -3	0.211507697723475\\
  -2	0.211514180454602\\
  -1	0.21152192506903\\
  0	0.211580557904903\\
  1	0.211552397440804\\
  };
\label{sf10_nc_it1}

\addplot [color=groseille, line width=1.5pt]
  table[row sep=crcr]{%
  -29	0.287150325329248\\
  -28	0.286062589302398\\
  -27	0.283873347478745\\
  -26	0.278110146948296\\
  -25	0.266739253738472\\
  -24	0.248896139090066\\
  -23	0.22453705692926\\
  -22	0.195237895758092\\
  -21	0.162086757895543\\
  -20	0.128899388278024\\
  -19	0.0980014605417818\\
  -18	0.0723197187671545\\
  -17	0.0554509531693313\\
  -16	0.0472294714868763\\
  -15	0.0432942593291666\\
  -14	0.0408733796221078\\
  -13	0.038978895820208\\
  -12	0.0374463332090074\\
  -11	0.0361541086933865\\
  -10	0.0352491409400243\\
  -9	0.034441810949532\\
  -8	0.0338338089551138\\
  -7	0.0333125946580002\\
  -6	0.0329447500587047\\
  -5	0.0326412821154453\\
  -4	0.0324084309546637\\
  -3	0.0321270741156559\\
  -2	0.0320109277286709\\
  -1	0.0319049088963457\\
  0	0.0318149540943622\\
  1	0.0317953320876735\\
  };
\label{sf10_nc_it2}

\end{axis}

\node [draw,fill=white,inner sep=2pt,at={(main.north east)},yshift=-15pt,xshift=2pt,anchor=south east ,]{
    \scriptsize
	\setlength{\tabcolsep}{2pt}
	\begin{tabular}{r|cccc}
        SF	& $7$ &$8$&$9$&$10$  \\
        \hline	\vspace{-6pt}
        & & \\
        w/o SFO comp. &\ref{sf7_nc_it1} &\ref{sf8_nc_it1}&\ref{sf9_nc_it1} &\ref{sf10_nc_it1} \\
        w/ SFO est. \& comp. &\ref{sf7_nc_it2} &\ref{sf8_nc_it2}&\ref{sf9_nc_it2} &\ref{sf10_nc_it2}\\
    \end{tabular}
};
\end{tikzpicture}%
  \caption{RMSE of fractional STO estimate with a frequency offset $\gamma=\SI{40}{ppm}$ for different spreading factors}
  \label{fig:sto_frac_err}
\end{figure}
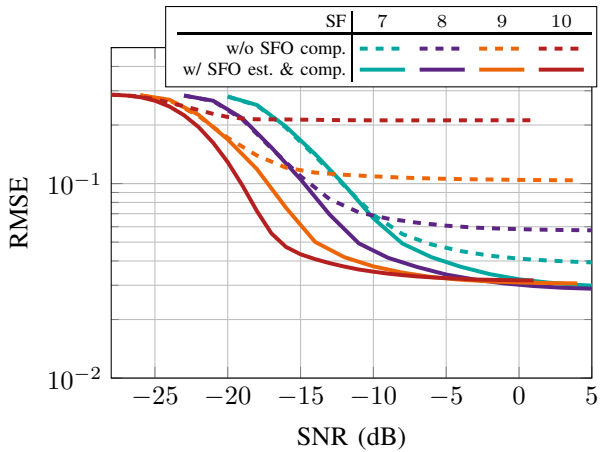

\figurename~\ref{fig:cfo_int_err} presents the root mean square error~(RMSE) of the integer CFO estimate for a reference clock frequency offset of $\gamma=\SI{40}{ppm}$.
Without the SFO compensation, we observe two different signal-to-noise ratios~(SNR) operating regimes (e.g., below and above \SI{-13}{\decibel} SNR for SF $10$).
Below this threshold, the estimation error is dominated by the AWGN, which causes integer STO estimation errors in the range $[\nicefrac{-2^\text{SF}}{2},\nicefrac{2^\text{SF}}{2})$.
Above this threshold, the error rate only slowly decreases with the SNR.
As discussed in Section~\ref{sec:sfo}, the dominant error is caused by poor fractional estimates of CFO and STO which cause integer STO estimation errors only in the range $[-1,1]$.
With our proposed synchronization scheme, only the error region dominated by noise remains.
We should note that the error estimation of the integer STO follows the same error curves since both integer offsets are estimated together.
\figurename~\ref{fig:sto_frac_err} shows the RMSE of the fractional STO estimate for $\gamma=\SI{40}{ppm}$.

We observe that without considering the SFO in the preamble, the estimation error exhibits a floor, which worsens as the spreading factor increases.
While the error floor is close to the limit of the estimation method for SF $7$, the SF $10$ already leads to an error floor close to a random guess.
\textcolor{black}{While an error floor remains after the SFO compensation, this floor is much lower than the resolution required to perform a realignment of the oversampled signal (i.e., $2B \,{\leq}\, f_s\,{\leq}\, 8B$ in practice).}
\subsection{Symbol Error Rate}
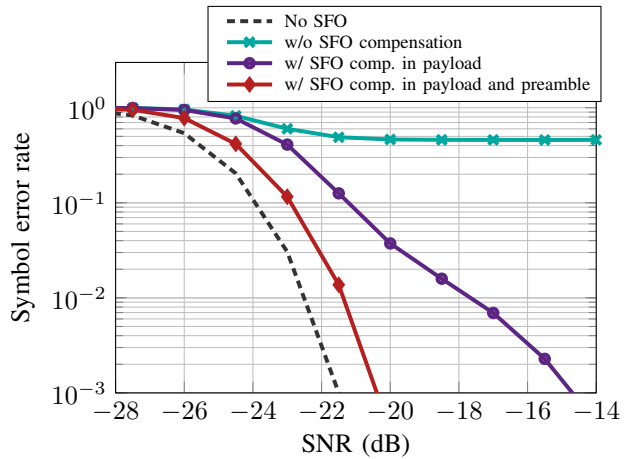
\begin{figure}
  \centering
  \begin{tikzpicture}
\begin{axis}[%
name=main,
width=.9\linewidth,
height=0.675\linewidth,
ymode=log,
xmin=-28,
xmax=-14,
ymin=1e-3,
ymax=3,
xlabel={SNR (dB)},
ylabel={Symbol error rate},
title style={at={(0.5,-0.3)}, anchor=north},
xticklabel shift={0pt},
xlabel style={yshift=0pt, },
xlabel style={yshift=3pt,},
axis background/.style={fill=white},
xmajorgrids,
xminorgrids,
ymajorgrids,
yminorgrids,
]
\addplot [densely dashed, color=black!80!white, line width=1.5pt]
  table[row sep=crcr]{%
-35	0.999747395833333\\
-33.5	0.9996171875\\
-32	0.998979166666666\\
-30.5	0.994106770833333\\
-29	0.962473958333333\\
-27.5	0.829786458333333\\
-26	0.539716145833333\\
-24.5	0.203997395833333\\
-23	0.0304895833333333\\
-21.5	0.000997395833333334\\
-20	7.8125e-06\\
-18.5	0\\
-17	0\\
-15.5	0\\
-14	0\\
-12.5	0\\
-11	0\\
};
\label{no_sfo}

\addplot [color=leman,mark=x,mark size=2.5pt, line width=1.5pt]
  table[row sep=crcr]{%
-35	0.999756944444445\\
-33.5	0.999774305555556\\
-32	0.999652777777778\\
-30.5	0.999739583333334\\
-29	0.999236111111111\\
-27.5	0.995173611111111\\
-26	0.960243055555556\\
-24.5	0.821996527777778\\
-23	0.601197916666667\\
-21.5	0.489548611111111\\
-20	0.464097222222222\\
-18.5	0.459774305555556\\
-17	0.459114583333333\\
-15.5	0.458489583333333\\
-14	0.459305555555556\\
-12.5	0.457934027777778\\
-11	0.458940972222222\\
};
\label{sf12_it1}

\addplot [color=zinzolin,mark=o,mark size=1.6pt, line width=1.5pt]
  table[row sep=crcr]{%
-35	0.999708333333333\\
-33.5	0.99971875\\
-32	0.99965625\\
-30.5	0.999354166666667\\
-29	0.998479166666667\\
-27.5	0.991234375\\
-26	0.94784375\\
-24.5	0.771994791666667\\
-23	0.409588541666667\\
-21.5	0.125760416666667\\
-20	0.037390625\\
-18.5	0.0158958333333333\\
-17	0.0069375\\
-15.5	0.00228125\\
-14	0.000489583333333333\\
-12.5	0.00015625\\
-11	5.20833333333333e-05\\
};

\label{sf12_it1_sfo}

\addplot [color=groseille,mark=diamond, line width=1.5pt]
  table[row sep=crcr]{%
-35	0.9996875\\
-33.5	0.999826388888889\\
-32	0.999704861111111\\
-30.5	0.999236111111111\\
-29	0.994375\\
-27.5	0.953524305555555\\
-26	0.778715277777778\\
-24.5	0.416666666666667\\
-23	0.115850694444444\\
-21.5	0.0137152777777778\\
-20	0.000399305555555556\\
-18.5	0\\
-17	0\\
-15.5	0\\
-14	0\\
-12.5	0\\
-11	0\\
};
\label{sf12_it2_sfo}

\end{axis}
\node [draw,fill=white,inner sep=2pt,at={(main.north east)},yshift=-15pt,xshift=2pt,anchor=south east,]{
    \scriptsize
	\setlength{\tabcolsep}{2pt}
	\begin{tabular}{l l}
    \ref{no_sfo} &  No SFO  \\
    \ref{sf12_it1} & w/o SFO compensation  \\
    \ref{sf12_it1_sfo} & w/ SFO comp. in payload  \\
    \ref{sf12_it2_sfo} &  w/ SFO comp. in payload and preamble  \\
    \end{tabular}
};

\end{tikzpicture}%
  \vspace{-5pt}
  \caption{Symbol error rate for SF $12$, $\gamma\,{=}\,\SI{32}{ppm}$, and 8 payload symbols per frame for different SFO compensations}
  \label{fig:sfo_ser}
\end{figure}

We now look at the symbol error rate~(SER) obtained for a spreading factor $12$ and a clock offset of $\gamma=\SI{32}{ppm}$.
We transmit frames with a default preamble size, i.e., $12.25$ symbols, followed by $8$ data symbols which corresponds to the minimum payload size of a LoRa transmission.
\figurename~\ref{fig:sfo_ser} presents the SER for different SNR values.
Without any SFO compensation, the error rate quickly floors at \SI{46}{\percent} as we reach the systematic demodulation error after a few symbols.
By compensating the SFO only during the payload part of a frame, we do not reach an error floor anymore.
However, we observe a significant performance improvement by additionally compensating the SFO in the preamble.
At a target SER of $10^{-3}$, we observe a gain of \SI{6}{\decibel} in SNR when using our proposed synchronization algorithm, compared to a simple SFO compensation during the data phase.
Finally, we can note that the performance of the synchronization algorithm is only \SI{1}{\decibel} away from the ideal case where no sampling frequency offset is present.
\section{Conclusion}
In this paper, we have shown that the presence of sampling frequency offset in the preamble of LoRa frames has a negative impact on the estimation of other impairments such as carrier frequency offset and sampling time offset.
We have briefly introduced state-of-the-art methods for time and frequency offset estimation for LoRa and detailed our proposed two-pass estimation process to improve the robustness of the synchronization algorithm against sampling frequency mismatch.
\textcolor{black}{We proposed to use the first estimation pass to obtain an estimate of the sampling frequency offset and then the second pass to refine the estimations of offsets that were degraded by the SFO.
In addition to removing any symbol error rate floor, the improved synchronization provides up to a \SI{6}{\decibel} gain in SNR at a target SER of $10^{-3}$ compared to a simple SFO compensation during the data phase.}
Finally, we have shown that the performance of the synchronization algorithm only exhibits a \SI{1}{\decibel} loss in SNR compared to an ideal scenario with no SFO.
\bibliographystyle{IEEEtran}
\bibliography{IEEEabrv,paper}
\end{document}